\documentclass[fleqn,10pt]{wlscirep}
\usepackage[utf8]{inputenc}
\usepackage[T1]{fontenc}
\newcommand{\be}{\begin{equation}}
\newcommand{\ee}{\end{equation}}

\newcommand{\bea}{\begin{eqnarray}}
\newcommand{\eea}{\end{eqnarray}}
\newcommand{\nn}{\nonumber}
\title{Dynamics of a V-type atom inside a deformed cavity field and in the presence of an external Microwave field}

\author[1]{E. Bolandhemmat}
\author[1,*]{F. Kheirandish}
\affil[1]{Department of Physics, University of Kurdistan, P.O.Box 66177-15175, Sanandaj, Iran}
\affil[*]{f.kheirandish@uok.ac.ir}

\begin{abstract}
\noindent In this article, we explore the interaction between a $V$-type atom inside a single mode deformed cavity field in the presence of an external microwave field. The Hamiltonian describing the system is derived from the standard Jaynes-Cummings model by deforming the field operators based on the Kerr-induced interaction. The total and reduced density matrices are obtained and the temporal evolution of nonclassical properties such as the Mandel Q parameter, quantum entanglement, and the position-momentum uncertainty relation (squeezing) of the field are examined. The impacts of coupling constant, generalized Kerr medium, and the intensity-dependent coupling function on the nonclassical indicators are thoroughly analyzed.
\end{abstract}
\begin{document}

\flushbottom
\maketitle
%
%
\thispagestyle{empty}

\section*{Introduction}

\label{Introduction}
\noindent Quantum electrodynamics (QED) stands at the forefront of quantum optics, significantly contributing to our understanding of the fundamental interactions between light and matter. Since the proposal and subsequent laboratory verification of the Purcell effect in the mid-20th century, cavity QED has provided an essential framework for exploring these interactions \cite{xue2016dynamics, bloembergen1948relaxation}. A typical cavity QED system consists of a high-finesse optical cavity that confines photons, coupled with an atomic subsystem. This setup allows for the confinement of light within a small volume, significantly enhancing the interaction between the atom and the electromagnetic field. The resulting system provides a versatile platform for studying a wide range of quantum phenomena, from quantum interference effects to the manipulation of quantum states and entanglement dynamics \cite{li2001full, wei2011broadband, sun2013electromagnetically, xue2015spontaneous, qiu2013spontaneous}.
The interaction between the confined electromagnetic field and atomic systems within the cavity leads to quantize field modes that significantly alter the properties of the atoms. This interaction is a cornerstone for understanding and controlling quantum phenomena, making cavity QED a pivotal field of study in contemporary physics \cite{armen2006low}. Further explorations in cavity QED have also delved into the potential for controlling and manipulating single photons, contributing to the development of quantum information processing technologies and the implementation of quantum networks. Cavity QED systems are not limited to simple two-level atoms; more complex atomic configurations, such as V-type three-level atoms, have also been extensively studied. In such systems, the interaction dynamics can be controlled using various external fields, such as lasers and microwave fields\cite{tang2015entanglement, cho2014quantum}. For a V-type three-level atom with two closely spaced upper energy levels, traditional methods involve using two different lasers finely tuned to specific wavelengths. However, combining a laser field with a microwave field offers an alternative approach to manipulating the atom-field interaction, providing additional degrees of control over the system\cite{huang2015drift, berman1997atom}.
The extension of atom-cavity interaction within the interacting Fock space has been documented. This model has been observed to exhibit Rabi oscillations\cite{muller2020dissipative, cai2021observation}. Multi-photon processes hold significant importance in atomic systems as they lead to a high correlation among emitted photons, culminating in the nonclassical behavior of the emitted light\cite{ghosh2011dynamics, rudolph1998multiphoton}. We have explored the dynamics of two-photon correlations produced by the interaction of a semiclassical three-level atom in configurations such as V, $\Lambda$ and $\Xi$, coupled with two classical external driving fields, under the rotating-wave approximation, and in the presence of atomic level decay terms. It has been noted that the correlation behavior is dependent on the configuration\cite{dhar2013controllable}. Adjusting the parameters of the driving fields and decay terms in various configurations results in differences in the characterization of the correlations.
The study of V-type three-level systems driven by an external microwave field within a cavity has revealed intriguing dynamics of population and quantum state control. Such systems allow for the acceleration of population transfer and the creation of maximal entanglement between atoms, crucial for practical applications in quantum information processing\cite{dhar2014mapping}. For instance, nonadiabatic schemes based on Berry’s transitionless quantum driving have been proposed to achieve faster population transfer in these systems \cite{lu2014shortcuts}. The Jaynes-Cummings Model (JCM) serves as a fundamental building block for understanding atom-field interactions\cite{shore1993jaynes}. Its simplicity and solvability make it an ideal tool for studying various quantum optical phenomena. The JCM describes the interaction of a two-level atom with a single mode of the electromagnetic field, capturing essential features such as spontaneous emission, Rabi oscillations, and atomic population inversion\cite{eberly1980periodic}. The model's predictions have been experimentally verified using Rydberg atoms in high-quality cavities, demonstrating the discrete and quantum nature of the interaction. The JCM has been instrumental in elucidating various nonclassical phenomena, including the periodic collapse and revival of Rabi oscillations, which highlight the discrete nature of photons. Extensions of the JCM have incorporated more complex scenarios, including multi-level atoms, multi-mode fields, and intensity-dependent coupling\cite{baghshahi2014entanglement, faghihi2014dynamics}. These extensions have led to the discovery of new quantum phenomena, such as quadrature squeezing, photon anti-bunching, and enhanced entanglement dynamics\cite{wang2019squeezing}. The inclusion of Kerr nonlinearity in the JCM has been particularly impactful, enabling the study of nonlinear optical effects and their applications in quantum non-demolition measurements and the generation of entangled macroscopic states\cite{chatterjee2019dynamics}.
In systems where the Kerr nonlinearity is incorporated into the JCM\cite{joshi1992dynamical}, the interaction between the atom and the field becomes more complex and rich in dynamics. For example, the Buck-Sukumar model, which includes intensity-dependent coupling, predicts the dependence of atom-field coupling on light intensity, leading to precise periodic behaviors in atomic population inversion and squeezing. For example, the Buck-Sukumar model demonstrated how atom-field coupling varies with light intensity\cite{buck1981exactly, sukumar1981multi}. Buek substantiated that with intensity-dependent coupling, precise periodicity in physical measurements, especially in atomic population inversion and squeezing, becomes observable when a two-level atom interacts with a single-mode field. An-fu and Zhi-wei discovered that the phase distribution is influenced by the intensity of the coherent field and the detuning parameter. Additionally, Fang et al\cite{fang2000super, an1994phase, fang1995properties}. explored how entropy and phase characteristics of the field are affected in the two-photon Jaynes-Cummings model under the influence of nonlinear interactions. Similarly, studies have shown that in the presence of Kerr nonlinearity, the phase distribution and entropy properties of the field are significantly altered, providing deeper insights into the nature of quantum correlations.

Here, we investigate the effects of nonlinearity function $f(\hat{n})$ and Kerr constant parameters on physical properties such as field entropy, squeezing, photon statistics, and Husimi distribution function \cite{almalki2024interaction}. For this purpose, we begin by examining the interaction between a V-type atom inside a deformed single mode cavity field and in the presence of an external microwave field. Our objective is to derive the explicit form of the state vector for the entire system by solving the time-dependent Schr\"{o}dinger equation, which we will detail in the following sections.\label{Introduction}
\noindent Quantum electrodynamics (QED) stands at the forefront of quantum optics, significantly contributing to our understanding of the fundamental interactions between light and matter. Since the proposal and subsequent laboratory verification of the Purcell effect in the mid-20th century, cavity QED has provided an essential framework for exploring these interactions \cite{xue2016dynamics, bloembergen1948relaxation}. A typical cavity QED system consists of a high-finesse optical cavity that confines photons, coupled with an atomic subsystem. This setup allows for the confinement of light within a small volume, significantly enhancing the interaction between the atom and the electromagnetic field. The resulting system provides a versatile platform for studying a wide range of quantum phenomena, from quantum interference effects to the manipulation of quantum states and entanglement dynamics \cite{li2001full, wei2011broadband, sun2013electromagnetically, xue2015spontaneous, qiu2013spontaneous}.
The interaction between the confined electromagnetic field and atomic systems within the cavity leads to quantize field modes that significantly alter the properties of the atoms. This interaction is a cornerstone for understanding and controlling quantum phenomena, making cavity QED a pivotal field of study in contemporary physics \cite{armen2006low}. Further explorations in cavity QED have also delved into the potential for controlling and manipulating single photons, contributing to the development of quantum information processing technologies and the implementation of quantum networks. Cavity QED systems are not limited to simple two-level atoms; more complex atomic configurations, such as V-type three-level atoms, have also been extensively studied. In such systems, the interaction dynamics can be controlled using various external fields, such as lasers and microwave fields\cite{tang2015entanglement, cho2014quantum}. For a V-type three-level atom with two closely spaced upper energy levels, traditional methods involve using two different lasers finely tuned to specific wavelengths. However, combining a laser field with a microwave field offers an alternative approach to manipulating the atom-field interaction, providing additional degrees of control over the system\cite{huang2015drift, berman1997atom}.
The extension of atom-cavity interaction within the interacting Fock space has been documented. This model has been observed to exhibit Rabi oscillations\cite{muller2020dissipative, cai2021observation}. Multi-photon processes hold significant importance in atomic systems as they lead to a high correlation among emitted photons, culminating in the nonclassical behavior of the emitted light\cite{ghosh2011dynamics, rudolph1998multiphoton}. We have explored the dynamics of two-photon correlations produced by the interaction of a semiclassical three-level atom in configurations such as V, $\Lambda$ and $\Xi$, coupled with two classical external driving fields, under the rotating-wave approximation, and in the presence of atomic level decay terms. It has been noted that the correlation behavior is dependent on the configuration\cite{dhar2013controllable}. Adjusting the parameters of the driving fields and decay terms in various configurations results in differences in the characterization of the correlations.
The study of V-type three-level systems driven by an external microwave field within a cavity has revealed intriguing dynamics of population and quantum state control. Such systems allow for the acceleration of population transfer and the creation of maximal entanglement between atoms, crucial for practical applications in quantum information processing\cite{dhar2014mapping}. For instance, nonadiabatic schemes based on Berry’s transitionless quantum driving have been proposed to achieve faster population transfer in these systems \cite{lu2014shortcuts}. The Jaynes-Cummings Model (JCM) serves as a fundamental building block for understanding atom-field interactions\cite{shore1993jaynes}. Its simplicity and solvability make it an ideal tool for studying various quantum optical phenomena. The JCM describes the interaction of a two-level atom with a single mode of the electromagnetic field, capturing essential features such as spontaneous emission, Rabi oscillations, and atomic population inversion\cite{eberly1980periodic}. The model's predictions have been experimentally verified using Rydberg atoms in high-quality cavities, demonstrating the discrete and quantum nature of the interaction. The JCM has been instrumental in elucidating various nonclassical phenomena, including the periodic collapse and revival of Rabi oscillations, which highlight the discrete nature of photons. Extensions of the JCM have incorporated more complex scenarios, including multi-level atoms, multi-mode fields, and intensity-dependent coupling\cite{baghshahi2014entanglement, faghihi2014dynamics}. These extensions have led to the discovery of new quantum phenomena, such as quadrature squeezing, photon anti-bunching, and enhanced entanglement dynamics\cite{wang2019squeezing}. The inclusion of Kerr nonlinearity in the JCM has been particularly impactful, enabling the study of nonlinear optical effects and their applications in quantum non-demolition measurements and the generation of entangled macroscopic states\cite{chatterjee2019dynamics}.
In systems where the Kerr nonlinearity is incorporated into the JCM\cite{joshi1992dynamical}, the interaction between the atom and the field becomes more complex and rich in dynamics. For example, the Buck-Sukumar model, which includes intensity-dependent coupling, predicts the dependence of atom-field coupling on light intensity, leading to precise periodic behaviors in atomic population inversion and squeezing. For example, the Buck-Sukumar model demonstrated how atom-field coupling varies with light intensity\cite{buck1981exactly, sukumar1981multi}. Buek substantiated that with intensity-dependent coupling, precise periodicity in physical measurements, especially in atomic population inversion and squeezing, becomes observable when a two-level atom interacts with a single-mode field. An-fu and Zhi-wei discovered that the phase distribution is influenced by the intensity of the coherent field and the detuning parameter. Additionally, Fang et al\cite{fang2000super, an1994phase, fang1995properties}. explored how entropy and phase characteristics of the field are affected in the two-photon Jaynes-Cummings model under the influence of nonlinear interactions. Similarly, studies have shown that in the presence of Kerr nonlinearity, the phase distribution and entropy properties of the field are significantly altered, providing deeper insights into the nature of quantum correlations.

Here, we investigate the effects of nonlinearity function $f(\hat{n})$ and Kerr constant parameters on physical properties such as field entropy, squeezing, photon statistics, and Husimi distribution function \cite{almalki2024interaction}. For this purpose, we begin by examining the interaction between a V-type atom inside a deformed single mode cavity field and in the presence of an external microwave field. Our objective is to derive the explicit form of the state vector for the entire system by solving the time-dependent Schr\"{o}dinger equation, which we will detail in the following sections.

\section*{Model and formalism}\label{sec1}
\begin{figure}[ht]
\centering
\includegraphics[scale=0.3]{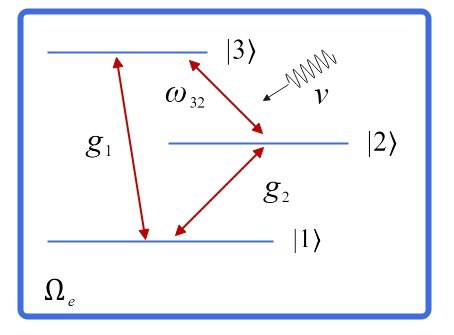}
\caption{(Color online) Schematic diagram of a three-level V-type atom confined inside
a single-mode cavity with coupling constant $ g_i$, atomic transition $|2\rangle\longleftrightarrow|3\rangle$ is driven by a microwave field with Rabi frequency $\Omega_e$ and carrier frequency $\nu$.}
\label{le}
\end{figure}

\noindent The present study offers a comprehensive analysis of the deformed model of a Three-level atomic (TLA) structure for a V-type atom inside a deformed single-mode cavity field whose dissipation is neglected, as depicted in  Fig. (\ref{le}). The two upper levels $|2\rangle$ and $|3\rangle$ are coupled to the lower level by the coupling constant $g_i$. The atomic transition is driven by a microwave classical field and for convenience, we assume that the system temperature is $0K$, where the dephasing rate of the quantum states can be neglected \cite{cohen1998atom}.
The model is governed by the Hamiltonian $\hat{H}$, consisting of an unperturbed part $H_0$ and an interaction part $H_1$. The unperturbed Hamiltonian, $H_0$, consists of two terms: $\Omega \hat{A}^\dag\hat{A}$, representing the photon field with frequency $\Omega$, and $\sum_{l}\omega_l\hat{\sigma}_{ll}$, denoting the energy levels of the atom with frequencies $\omega_l$, where $(\omega_3>\omega_2>\omega_l)$ \cite{scully1997quantum}. The interaction Hamiltonian, $H_1$, describes the coupling between the atomic levels and the photon field, where $g_1$ and $g_2$ are the coupling constants for transitions between the first and third levels, and the second and first levels, respectively and the transition $|3\rangle$$\rightarrow $$|2\rangle$ is forbidden in the electric-dipole approximation. These coupling constants are crucial as they determine the strength of interaction between the atom and the photon field, reflecting how effectively the photon field can induce transitions between different atomic states. Additionally, $\Omega_e$ represents the Rabi frequency associated with the transition between the second and third levels, which indicates the oscillation frequency of the atomic population between these states due to the external driving field with carrier frequency $\nu$.

Our study focuses on the eigenvalues and eigenvectors of this Hamiltonian, providing insights into the quantum dynamics and properties of the V-type atomic system under this deformed model. By solving the time-dependent Schr\"{o}dinger equation, the researchers can determine how the system evolves over time and how the deformations affect the behavior of the atom and photon field \cite{algarni2022parity, berrada2011construction, berrada2011maximal}.

The implications of this model are significant for understanding light-matter interactions in quantum optics, particularly in the precise control of quantum states, which is essential for applications in quantum information processing, such as the development of quantum computers and secure quantum communication systems, quantum jumps, quantum Zeno effect \cite{dhar2013controllable}. The ability to manipulate and control quantum states through tailored interactions opens new avenues for advancements in these cutting-edge technologies.
The deformation is incorporated into the atom-field system in the field properties are adjusted through an intensity-dependent coupling between the atom and the radiation field, meaning the coupling is no longer linear in relation to the field variables. The Hamiltonian that characterizes the dynamics of this quantum system within the Rotating Wave Approximation (RWA) is
\bea\label{HA}
\hat{H}=\underbrace{\Omega \hat{A}^\dag\hat{A}+\sum_{l}\omega_l\hat{\sigma}_{ll}}_{H_0}+\underbrace{g_1(\hat{A}\hat{\sigma}_{31}+\hat{A}^\dag\hat{\sigma}_{13})+g_2(\hat{A}\hat{\sigma}_{21}+\hat{A}^\dag\hat{\sigma}_{12})+\Omega_e(\hat{\sigma_{23}}+\hat{\sigma_{32}})}_{H_1},
\eea
where $\sigma_{ij}$ is the atomic density operator assigned by $\sigma_{ij}=|i\rangle\langle j|$, $\hat{A}^\dag=f(\hat{n})\hat{a}^\dag, \hat{A}=\hat{a}f(\hat{n}) $  are the f -deformed annihilation and creation operators constructed from the usual bosonic operators and satisfy the f -deformed bosonic commutation relations. It is important to note that choosing different nonlinearity functions results in distinct Hamiltonian systems, which can consequently produce a variety of physical outcomes. This variation in nonlinearity directly influences the dynamics and observable properties of the quantum system, leading to diverse phenomena and behaviors. The f-deformed bosonic commutation relations are defined by
\bea
&& [\hat{A},\hat{A}^\dag] = (\hat{n}+1)f^2(\hat{n}+1)-\hat{n}f^2(\hat{n})=k(\hat{n}),\nonumber\\
&& [\hat{A}^\dag,\hat {n}] = -\hat{A}^\dag,\nonumber\\
&& [\hat{A},\hat {n}] = \hat{A}.
\eea
By setting $f(\hat{n})=1$, the Hamiltonian Eq.~(\ref{HA}) transforms into the standard generalized Jaynes–Cummings model with classical microwave field and commutation
relations known as Heisenberg-Weyl algebra defined by non-deformed annihilation an creation operators $\hat{a},\hat{a}^\dag$.

The Hamiltonian in the interaction picture, $\hat{H}_{int}$ is given by
\bea
\hat{H}_{int}=e^{i\hat{H}_0t}\hat{H}_1e^{-i\hat{H}_0t},
\eea
where $\hat{H}_0$ is defined in Eq.~(\ref{HA}). By making use of the Baker-Campbell-Hausdorff formula, the interaction Hamiltonian $\hat{H}_{int}$ can be approximated as
\bea\label{Hau}
\hat{H}_{int}=\hat{H}_1+it[\hat{H}_0,\hat{H}_1]+\frac{it}{2!}[\hat{H}_0,[\hat{H}_0,\hat{H}_1]]+...,
\eea
therefore,
\bea\label{Hint}
\hat{H}_{int}=g_1(e^{-ith(\hat{n})}\hat{A}\hat{\sigma}_{31}+\hat{A}^\dag e^{ith(\hat{n})}\hat{\sigma}_{13})+g_2(e^{-its(\hat{n})}\hat{A}\hat{\sigma}_{21}+\hat{A}^\dag e^{its(\hat{n})}\hat{\sigma}_{12})+\Omega_e( e^{it(\omega_2-\omega_3)}\sigma_{23}+e^{it(\omega_3-\omega_2)}\sigma_{32}),
\eea
where for convenience we defined the following equations
\bea\label{hs}
h(\hat{n})=\Omega k(\hat{n})-(\omega_3-\omega_1),\nonumber\\
s(\hat{n})=\Omega k(\hat{n})-(\omega_2-\omega_1).
\eea
In quantum mechanics, the wave function of a physical system encapsulates all the possible information about the state of the system. By solving  the schr\"{o}dinger  equation of motion, one can determine the time evolution of this wave function, thereby predicting the probabilities of various outcomes for measurements on the system. This foundational equation describes how the quantum state changes over time under the influence of a given Hamiltonian, providing insights into the behavior and properties of the quantum system. The solutions to the Schr\"{o}dinger equation enable us to understand phenomena such as energy quantization, superposition, and entanglement, which are central to quantum theory.
\bea\label{Sch}
i\frac{\partial|\psi\rangle}{\partial t}=\hat{H}|\psi\rangle.
\eea
we consider the wave function $|\psi\rangle$ represented as a superposition of states in invariant subspaces denoted by quantum number $n$. The $n$-sector invariant space is spanned by the states $|1,n+1\rangle$, $|2,n\rangle$ and $|3,n\rangle$, and the corresponding amplitudes are denoted as $c_1(n+1,t), c_2(n,t)$ and $c_3(n,t)$, respectively.
The expressions $e^{-ih(\hat{n})t}$ and $e^{-is(\hat{n})t}$ describe the phase evolution due to the interaction Hamiltonian, which are functions of the number operator $\hat{n}$, understanding this detailed phase evolution is essential for analyzing phenomena such as stimulated emission, absorption, and nonlinear effects in quantum optics and quantum information processing. The total state of the system can be written as
\bea\label{si}
|\psi(t)\rangle=\sum_{n=0}^{\infty}(c_1(n+1,t)|1,n+1\rangle+c_2(n,t)|2,n\rangle+c_3(n,t)|3,n\rangle).
\eea
By inserting Eqs.~(\ref{si}) into Schr\"{o}dinger equation Eq. (\ref{Sch}), we obtain
\begin{eqnarray}\label{c123}
    &&i\dot{c_1}(n+1,t)=\underbrace{g_1 f(n+1)\sqrt{n+1}}_{v_1(n)}\,e^{ith(n)}c_3(n,t)+\underbrace{g_2 f(n+1)\sqrt{n+1}}_{v_2 (n)}\,e^{its(n)}c_2(n,t),\nn\\
    &&i\dot{c_2}(n,t)=g_2 f(n+1)\sqrt{n+1}\,e^{-its(n)}c_1(n+1,t)+\Omega_e e^{it(\omega_2-\omega_3)}c_3(n,t),\nn\\
   &&i\dot{c_3}(n,t)=g_1 f(n+1)\sqrt{n+1}e^{-ith(n)}c_1(n+1,t)+\Omega_e e^{-it(\omega_2-\omega_3)}c_2(n,t).
\end{eqnarray}
From Eq.~(\ref{hs}), we have $h(n)=s(n)-\nu$, where $\nu=\omega_{32}$. From Eqs.~(\ref{c123}) we have
\begin{eqnarray}\label{abc1}
    &&\dot{c_1}(n+1,t)=-iv_1(n)e^{ith(n)}c_3(n,t)-iv_2(n)e^{its(n)}c_2(n,t),\nn\\
    &&\dot{c_2}(n,t)=-iv_2(n)e^{-its(n)}c_1(n+1,t)-i\Omega_e c_3(n,t)e^{-i\nu t},\nn\\
    &&\dot{c_3}(n,t)=-iv_1(n)e^{-ith(n)}c_1(n+1,t)-i\Omega_e c_2(n,t)e^{i\nu t}.
\end{eqnarray}
By taking the Laplace transform of both sides of Eqs.~(\ref{abc1}), we obtain
\begin{eqnarray}
    \label{abc11}
    &&s\tilde{c_1}(n+1,s)-c_1(n+1,0)=-iv_1(n)\tilde c_3(n,s-ih(n))-iv_2(n)\tilde c_2(n,s-is(n)),\nn\\
    &&s\tilde{c_2}(n,s)-c_2(n,0)=-iv_2(n)\tilde{c_1}(n+1,s+is(n))-i\Omega_e \tilde{c_3}(n,s+i\nu), \;\;\;\;   \,\,s\longrightarrow s-is(n).\nn\\
    &&s\tilde{c_3}(n,s)-c_3(n,0)=-iv_1(n)\tilde{c_1}(n+1,s+ih(n))-i\Omega_e \tilde{c_2}(n,s-i\nu).  \;\;\;\;  \,\,s\longrightarrow s-ih(n),\nn\\
\end{eqnarray}
By inserting the above transformations we will have:
\begin{eqnarray}
    \label{abc11}
   && s\tilde{c_1}(n+1,s)-c_1(n+1,0)=-iv_1(n)\tilde c_3(n,s-ih(n))-iv_2(n)\tilde c_2(n,s-is(n)),\nn\\
    && (s-is(n))\tilde{c_2}(n,s-is(n))-c_2(n,0)=-iv_2(n)\tilde{c_1}(n+1,s)-i\Omega_e \tilde{c_3}(n,\underbrace{s-is(n)+i\nu}_{(s-ih(n))}), \;\;\;\;   \,\,s\longrightarrow s-is(n).\nn\\
   && (s-ih(n))\tilde{c_3}(n,s)-c_3(n,0)=-iv_1(n)\tilde{c_1}(n+1,s)-i\Omega_e \tilde{c_2}(n,\underbrace{s-ih(n)+i\nu}_{(s-is(n))}).  \;\;\;\;  \,\,s\longrightarrow s-ih(n),\nn\\
\end{eqnarray}
which can be rewritten in matrix form as
\bea
\underbrace{\begin{pmatrix}
  s & iv_2(n) & iv_1(n) \\
  iv_2(n) & s-is(n) & i\Omega_e \\
  iv_1(n) & i\Omega_e & s-ih(n)
\end{pmatrix}}_{M(s)}
\begin{pmatrix}
  \tilde{c_1}(n+1,s) \\
  \tilde{c_2}(n,s-is(n)) \\
  \tilde{c_3}(n,s-ih(n))
\end{pmatrix}
=
\begin{pmatrix}
  c_1(n+1,0) \\
  c_2(n,0) \\
   c_3(n,0)
\end{pmatrix}\label{matrix}.
\eea
Therefore,
\bea\label{ABC}
\begin{array}{lll}
\tilde{c}_1(n+1,s)=\sum\limits_{j=1}^{3}[M^{-1}(s)]_{1j}\,c_j(0),\\
\tilde{c}_2(n,s-is(n))=\sum\limits_{j=1}^{3}[M^{-1}(s)]_{2j}\,c_j(0,\\
\tilde{c}_3(n,s-ih(n))=\sum\limits_{j=1}^{3}[M^{-1}(s)]_{3j}\,c_j(0).\\
\end{array}
\eea
Now by taking the inverse Laplace transform of both sides of Eqs.~(\ref{ABC}), we find
\bea
\begin{pmatrix}
 c_1(n+1,t) \\
 e^{its(n)}c_2(n,t) \\
  e^{ith(n)}c_3(n,t)
\end{pmatrix}
=\mathcal{L}^{-1} ([M^{-1}(s)]
\begin{pmatrix}
  c_1(n+1,0) \\
  c_2(n,0) \\
   c_3(n,0)
\end{pmatrix}\label{matrix}.
\eea
Finally
\bea\label{ABC1}
\begin{array}{lll}
c_1(n+1,t)=\sum\limits_{j=1}^{3}\mathcal{L}^{-1} ([M^{-1}(s)]_{1j})(t)\,c_j(0),\\
c_2(n,t)= e^{-its(n)}\sum\limits_{j=1}^{3}\mathcal{L}^{-1} ([M^{-1}(s)]_{2j})(t)\,c_j(0),\\
c_3(n,t)= e^{-ith(n)}\sum\limits_{j=1}^{3}\mathcal{L}^{-1} ([M^{-1}(s)]_{3j})(t)\,c_j(0).
\end{array}
\eea
If we assume $\mathcal{L}^{-1} ([M^{-1}(s)]_{ij})=\sum\limits_{j=1}^{3}A_{ij}e^{\alpha_jt}$, ($i=1,2,3$), then the explicit formulas for the time-dependent amplitudes in the $n$-sector invariant space are
\bea\label{ABC2}
&&c_1(n+1,t)=\sum_{j=1}^{3}A_{1j}e^{\alpha_jt}c_j(0),\nn\\
&&c_2(n,t)=\sum_{j=1}^{3}A_{2j}e^{(\alpha_j-is(n))t}c_j(0),\nn\\
&&c_3(n,t)=\sum_{j=1}^{3}A_{3j}e^{(\alpha_j-ih(n))t}c_j(0),
\eea
where $A_{1j}$, $A_{2j}$ and $A_{3j}$’s can be evaluated using the initial conditions for the atom. We now assume that the atom initially enters the cavity in the excited state $|2\rangle$, that is the initial state of the system is $
|\psi(0)\rangle=|2\rangle\bigotimes|n\rangle$, where $|n\rangle$ represents a number state of the deformed cavity field. Therefore, $c_1(n+1,0)=0$, $c_2(n,0)=1$, and $c_3(n,0)=0$. Using these initial conditions, Eqs.~(\ref{ABC2}) can be rewritten in matrix form as
\bea
\begin{pmatrix}
  c_1(n+1,t) \\
 e^{its(n)}c_2(n,t) \\
  e^{ith(n)}c_3(n,t)
\end{pmatrix}
=\mathcal{L}^{-1}
\begin{pmatrix}
&&\frac{s^2-is (s(n)+s h(n) )+ \Omega_e^2 - s(n)h(n) }{\Theta(s)}&&-\frac{isv_2(n) + v_1(n)\Omega_e + h(n)v_2(n)}{\Theta(s)}&&-\frac{isv_1(n) + v_2(n)\Omega_e + s(n)v_1(n)}{\Theta(s)} \\
  &&-\frac{isv_2(n) + v_1(n)\Omega_e + h(n)v_2(n)}{\Theta(s)}&&\frac{s^2-ih(n)s+ v^2_1(n)}{\Theta(s)}&&-\frac{is\Omega_e + v_1(n)v_2(n)}{\Theta(s)}\\
 && -\frac{isv_1(n)+ v_2(n)\Omega_e + s(n)v_1(n)}{\Theta(s)}&&-\frac{is\Omega_e + v_1(n)v_2(n)}{\Theta(s)}&&\frac{s^2-is s(n) + v^2_2(n)}{\Theta(s)}
\end{pmatrix}
\begin{pmatrix}
  0 \\
  1\\
  0
\end{pmatrix},\nn\\
\eea
or equivalently,
\bea
\begin{pmatrix}
  c_1(n+1,t) \\
 e^{its(n)}c_2(n,t) \\
  e^{ith(n)}c_3(n,t)
\end{pmatrix}
=\mathcal{L}^{-1}
\begin{pmatrix}
  -\frac{isv_2(n) + v_1(n)\Omega_e + h(n)v_2(n)}{\Theta(s)}\\
 \frac{ s^2-ish(n)+v^2_1(n)}{\Theta(s)}\\
  - \frac{is\Omega_e + v_1(n)v_2(n)}{\Theta(s)}\nn\\
\end{pmatrix}\\
\eea
To proceed, we need the roots of the cubic equation $\Theta(s)=s^3-is^2(h(n)+s(n))+s(\Omega_e^2+v_1^2(n)+v_2^2(n)-s(n)h(n))-2i\Omega_e v_1(n)v_2(n)-iv^2_1(n)s(n)-iv^2_2(n)h(n)$. If we denote the roots by $\alpha_1(n),\,\alpha_2(n),\,$ and $\alpha_3(n)$, then we can write
\bea
c_1(n+1,t)&=&\frac{h(n)v_2(n)+v_1(n)\Omega_e+i\alpha_1(n)v_2(n)}{(\alpha_1(n)-\alpha_2(n))(\alpha_1(n)-\alpha_3(n))}e^{\alpha_1(n)t}+\frac{h(n)v_2(n)+v_1(n)\Omega_e+i\alpha_2(n)v_2(n)}{(\alpha_1(n)-\alpha_2(n))(\alpha_2(n)-\alpha_3(n))}e^{\alpha_2(n)t}\nn\\
&+&\frac{h(n)v_2(n)+v_1(n)\Omega_e+i\alpha_3(n)v_2(n)}{(\alpha_1(n)-\alpha_3(n))(\alpha_3(n)-\alpha_2(n))}e^{\alpha_3(n)t},\\
c_2(n,t)&=&\frac{\alpha^2_1-ih(n)\alpha_1(n)+v^2_1}{(\alpha_1(n)-\alpha_2(n))(\alpha_1(n)-\alpha_3(n))}e^{(\alpha_1(n)-is(n))t}+\frac{-\alpha^2_2+ih(n)\alpha_2(n)-v_1^2(n)}{(\alpha_1(n)-\alpha_2(n))(\alpha_2(n)-\alpha_3(n))}e^{(\alpha_2(n)-is(n))t}\nn\\
&+&\frac{\alpha^2_3-ih(n)\alpha_3(n)+v_1^2(n)}{(\alpha_1(n)-\alpha_3(n))(\alpha_3(n)-\alpha_2(n))}e^{(\alpha_3(n)-is(n))t},\\
c_3(n,t)&=&\frac{-i\Omega_e\alpha_1(n)-v_1(n)v_2(n)}{(\alpha_1(n)-\alpha_2(n))(\alpha_1(n)-\alpha_3(n))}e^{(\alpha_1(n)-ih(n))t}+\frac{i\Omega_e\alpha_2(n)+v_1(n)v_2(n)}{(\alpha_1(n)-\alpha_2(n))(\alpha_2(n)-\alpha_3(n))}e^{(\alpha_2(n)-ih(n))t}\nn\\
&+&\frac{-i\Omega_e\alpha_3(n)-v_1(n)v_2(n)}{(\alpha_1(n)-\alpha_3(n))(\alpha_3(n)-\alpha_2(n))}e^{(\alpha_3(n)-ih(n))t}.
\eea
In the Fig.~(\ref{pro}), we have plotted the probability distribution function for different parameters($\Omega_e, \chi$ and $g_{i}$) and analyzed the impact of applying an external microwave field between the two upper states. An illustrative example is when we select $f(\hat{n})=\sqrt{1+k\hat{n}^2}$, where $k$ is a positive constant $(k\geq 0)$. This form is similar to the Kerr-induced interaction with $\chi=k\Omega$. This addition leads to population transfer between the two upper states, resulting in a series of intriguing phenomena. Firstly, the population dynamics of the two upper levels no longer follow the same pattern over time, as the populations of the two excited states are perturbed by the external field, the deformation of Kerr-induced interaction and the coupling effects.\\
To elucidate the role of the Kerr effect and the coupling intensity more explicitly, the results are depicted in the following figures. It can be observed when $\chi=0$ the frequency of the slow oscillations increases gradually as the coupling and Rabi frequency parameters rises. In cases with $f(\hat{n})=1$, as shown in the top row (a, b, c) of Figs. (\ref{pro}), the influence of the microwave field is more pronounced, with the profile of the slow oscillations being more distinct. Conversely, for larger values of the intensity-dependent coupling, Rabi frequency of the external
field and coupling ($g_i$), the effect of the cavity photon becomes more dominant.

Additionally, the amplitudes of the two upper levels increase significantly as the parameters
$\Omega_e$, $\chi$ and $g_i$ grows. This suggests that the population transfer between the upper levels can be more complete with higher intensities of the external field and Kerr constant, and the effect of the cavity photon becomes more dominant in cases where the parameter $g$ is larger. These findings highlight the complex interplay between the microwave field and the cavity photon field, demonstrating how varying these parameters can lead to different dynamic behaviors in the quantum system.\\
The mechanism of the cavity photon and microwave field under the influence of a f-deformed function on a V-type three-level atom has been elucidated through the above theoretical analysis and numerical results. Conducting experiments using this scheme is also feasible, as the technique of irradiating a nano-cavity containing atoms with a microwave field is well established and commonly used. Therefore, our theoretical model and results can provide a good reference for experimental details.
\begin{table}[ht!]
\centering
\begin{tabular}{ccc}
    \includegraphics[width=0.3\textwidth]{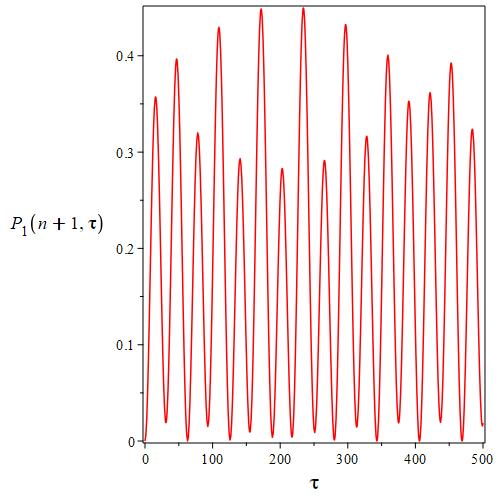} &
    \includegraphics[width=0.3\textwidth]{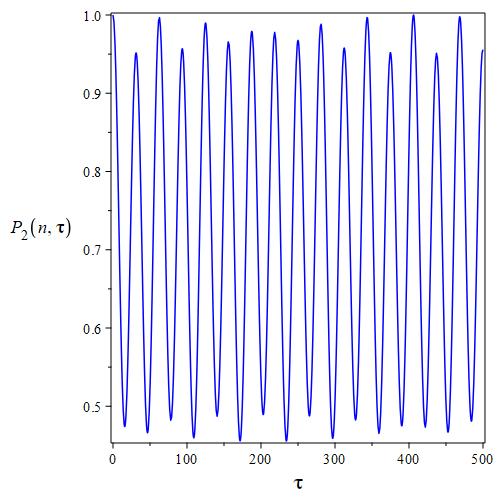} &
    \includegraphics[width=0.3\textwidth]{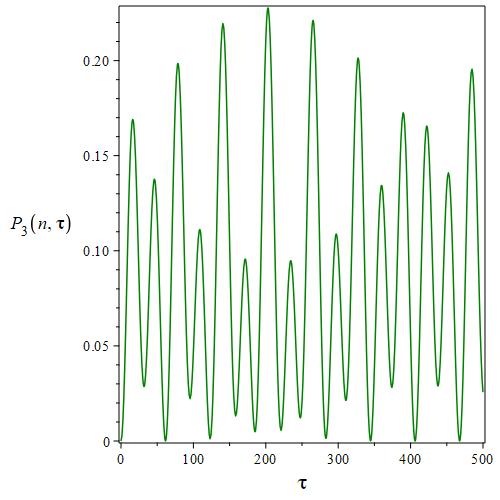} \\
    \text{a} &\label{pro1}
    \text{b} &\label{pro2}
    \text{c} \\ \label{pro3}
    \includegraphics[width=0.3\textwidth]{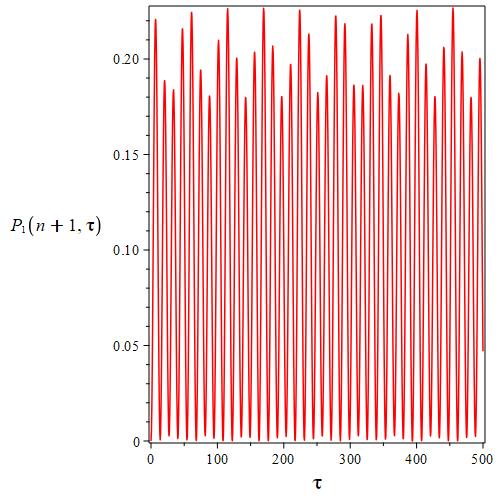} &
    \includegraphics[width=0.3\textwidth]{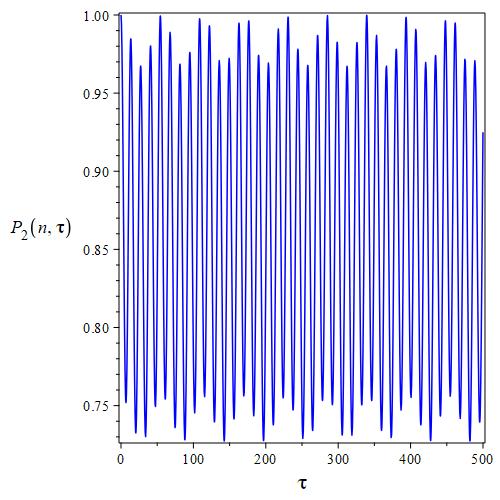} &
    \includegraphics[width=0.3\textwidth]{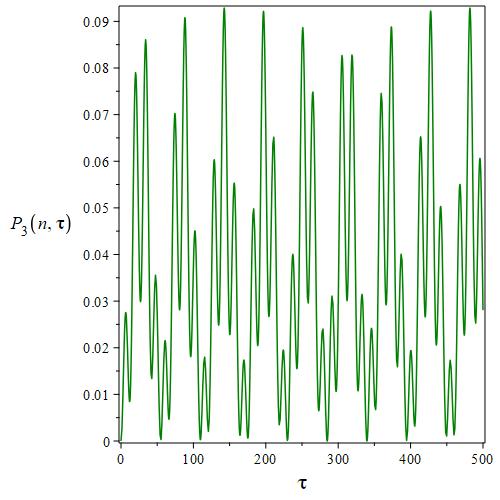} \\
    \text{d} &
    \text{e} &
    \text{f} \\
    \includegraphics[width=0.3\textwidth]{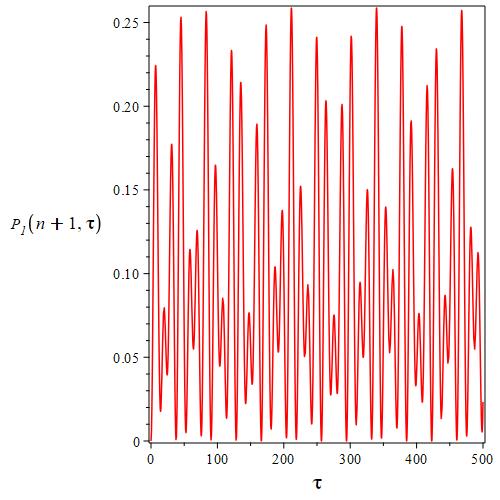} &
    \includegraphics[width=0.3\textwidth]{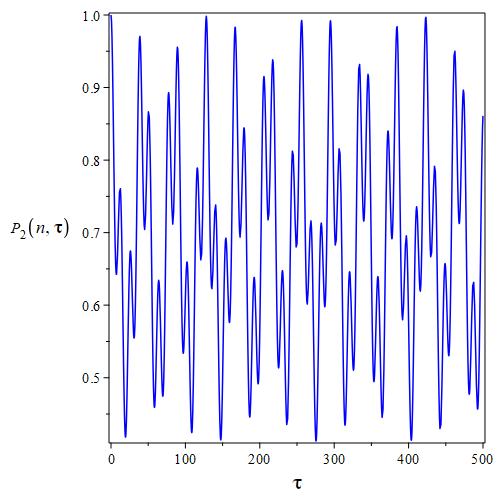} &
    \includegraphics[width=0.3\textwidth]{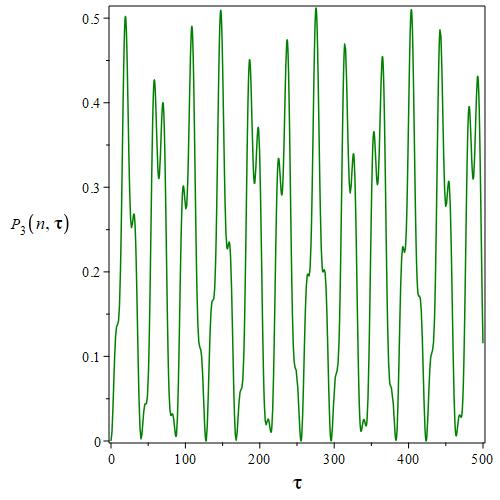} \\
    \text{g} &
    \text{h} &
    \text{i} \\
\end{tabular}
 \caption{(Color online)  The evolution of probabilities for the parameters $\Omega=0.2$, $\omega_3=0.5, \omega_2=0.4, \omega_1=0.3$, $n=1$, the intensity-dependent coupling $f(\hat{n})=\sqrt{1+\chi\hat{n}^2}$ and the $\tau=\Omega t$.
  The upper row. (a,b,c) corresponds to the $\Omega_e=0.04, g_1=0.04, g_2=0.06$ and $\chi=0$, the middle row. (d,e,f) corresponds to the $\Omega_e=0.04, g_1=0.06, g_2=0.08$ and $\chi=0.2$ and the bottom row. (g,h,i) corresponds to the $\Omega_e=0.08, g_1=0.06, g_2=0.08$ and $\chi=0.2$.}
 \label{pro}
\end{table}
Based on the TLA-field wave function $|\psi(t)\rangle$, we can extract the time-dependent properties of various quantum phenomena that are associated with the proposed system. Here, we consider the population inversion, quantum coherence and von Neumann entropy, which depend on the the density matrix elements of $\rho_{A}(T)$.
\bea
\rho_{A}(T)=Tr_{F}|\psi(t)\rangle\langle \psi(t)|=\sum_{j=0}^{3}\sum_{l=0}^{3}\rho_{jl}(T)|j\rangle\langle l|,
\eea
where we have used the subscript $A (F)$ to denote the atom (field).
\section*{Quantum metrics and computational outcomes}\label{sec2}
\noindent In order to display the deformation of the field on the quantum quantifiers, we show the temporal evolution of the atomic inversion, second-order correlation function, atomic entropy.
\subsection{Population inversion}
\noindent In the view of quantum optics and information, a crucial quantity is the population inversion. This measure helps identify the moments of collapse and revival, which play a key role in defining the durations of separable and maximally entangled states \cite{olaya2004scheme, yang2004interactions, you2018quantum}. The atomic population inversion of $\rho_A(T)$ is defined by
\bea
W=\rho_{11}-\rho_{33}.
\eea
\begin{table}[ht!]
\centering
\begin{tabular}{ccc}
    \includegraphics[width=0.3\textwidth]{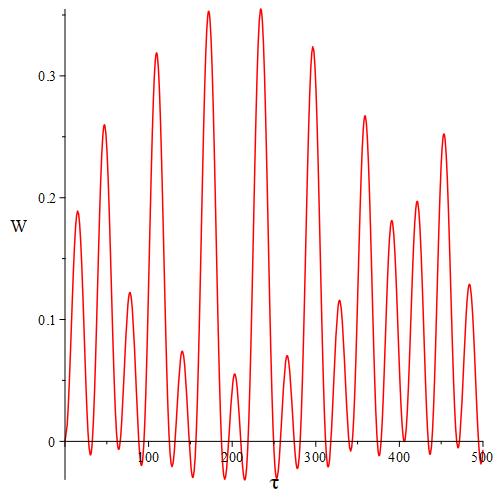} &
    \includegraphics[width=0.3\textwidth]{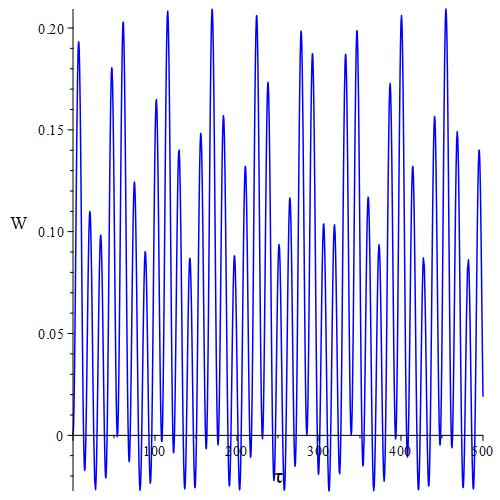} &
    \includegraphics[width=0.3\textwidth]{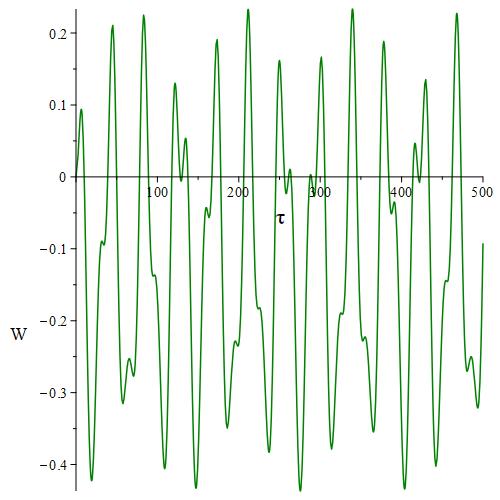} \\
     \text{a}\label{fig:w1}&
    \text{b}\label{fig:w2}&
    \text{c}\label{fig:w3}\\
\end{tabular}
  \caption{(Color online) The evolution of atomic inversion for the parameters $\Omega=0.2$, $\omega_3=0.5, \omega_2=0.4, \omega_1=0.3$, $n=1$, the intensity-dependent coupling $f(\hat{n})=\sqrt{1+\chi\hat{n}^2}$ and the $\tau=\Omega t$.
  The Fig. (a) corresponds to the $\Omega_e=0.04, g_1=0.04, g_2=0.06$ and $\chi=0$, the Fig. (b) corresponds to the $\Omega_e=0.04, g_1=0.06, g_2=0.08$ and $\chi=0.2$ and the Fig. (c) corresponds to the $\Omega_e=0.08, g_1=0.06, g_2=0.08$ and $\chi=0.2$.}\label{fig:inversion}
\end{table}
In Fig. (\ref{fig:inversion}), the function $W$ randomly oscillates between the excited state and the ground state during the considered interaction period.  We observe that the atomic inversion, when the coupling parameters, Rabi frequency, and $\chi$ are small, exhibits negligible oscillations with phenomena of revival and collapse, leading to behavior where the measurement of atomic inversion is accompanied by a reduction in oscillation amplitudes. From these results, we conclude that the coupling parameter can enhance the oscillation range in atomic inversion and that the presence of the $\Omega_e=0.08$, $\chi=0.2$ can increase the the number of oscillation range in atomic inversion.
\section*{Nonclassical effects}\label{sec3}
\noindent The second-order correlation function $g^2(\tau)$, is a measure of the intensity correlations in a quantum or classical field and plays a fundamental role in the field of quantum optics. These correlations provide critical insights into the statistical properties of light and the underlying quantum states of electromagnetic fields. The second-order correlation function measures the likelihood of detecting two photons separated by a time interval $\tau$, compared to the likelihood expected for a completely random (coherent) light source. This function is essential for distinguishing between classical and quantum states of light, as it can reveal phenomena such as photon bunching and anti-bunching \cite{gerry2023introductory}.
 It is defined as
 \bea
g^2(\tau)=\frac{\langle: I(t)I(t+\tau):\rangle}{\langle I(t)\rangle^2},
 \eea
where $I(t)$  is the intensity of the light field at time $t$ and $: I(t)I(t+\tau):=\hat{A}^\dagger(t)\hat{A}^\dagger(t+\tau)\hat{A}(t+\tau)\hat{A}(t)$. The correlation $g^2(\tau)$ is proportional to the detection probability of one photon at the time $t$ and a second one at $(t+\tau)$
\bea\label{gg}
g^2(\tau)=\frac{\langle\hat{A}^\dagger(t)\hat{A}^\dagger(t+\tau)\hat{A}(t+\tau)\hat{A}(t)\rangle}{\langle\hat{A}^\dagger(t)\hat{A}(t)\rangle^2},
\eea
For our atomic system,
\bea
 \langle \hat{A}^\dagger\hat{A}\rangle=\sum_{n}[f^2(n+1)(n+1)|C_1(n+1,t)|^2+n(f^2(n)|C_2(n,t)|^2+f^2(n)|C_3(n,t)|^2)],\nn\\
  \langle (\hat{A}^\dagger\hat{A})^2\rangle=\sum_{n}[f^4(n+1)(n+1)^2|C_1(n+1,t)|^2+n^2f^4(n)(|C_2(n,t)|^2+|C_3(n,t)|^2)].
\eea
By substituting the nonlinear creation and annihilation operators in Eq. (\ref{gg}), we can examine the effects of nonlinearity in the second-order correlation function. In the context of quantum optics, where $\hat{A}$ and $\hat{A}^\dagger$ are the annihilation and creation operators, the second-order correlation function at zero delay $(\tau=0)$ is often used, and is given by:
\bea\label{gg}
g^2(0)&=&\frac{\langle \hat{A}^\dagger\hat{A}^\dagger\hat{A}\hat{A} \rangle}{\langle \hat{A}^\dagger\hat{A}\rangle^2},\nn\\
&=&\frac{(n+1)f^2(n+1)nf^2(n)|C_1(n+1,t)|^2+(n-1)f^2(n-1)nf^2(n)(|C_2(n,t)|^2+|C_3(n,t)|^2)}{((n+1)f^2(n+1)|C_1(n+1,t)|^2+nf^2(n)(|C_2(n,t)|^2+|C_3(n,t)|^2))^2}.
\eea
which is proportional to the detection probability of two photons in the same time. In the case where $g^2(0)<g^2(\tau)$, the detection probability of a second photon after a delay time $\tau$ decreases, indicating the phenomenon of photon anti-bunching. Anti-bunching is a signature of quantum light sources, such as single-photon emitters, where photons tend to be emitted one at a time rather than in pairs or groups. Conversely, when $g^2(0)>g^2(\tau)$, the detection probability of a second photon increases with the delay time, corresponding to the bunching effect. Photon bunching is characteristic of thermal or chaotic light sources, where photons are more likely to be detected in pairs or groups.

For $g^2(\tau)=1$ we observe the case of coherent states, such as those produced by an ideal laser, where the photon statistics follow a Poisson distribution. Coherent light does not exhibit any bunching or anti-bunching effects, as the photons are distributed randomly over time.
\begin{table}[ht!]
\centering
\begin{tabular}{ccc}
    \includegraphics[width=0.3\textwidth]{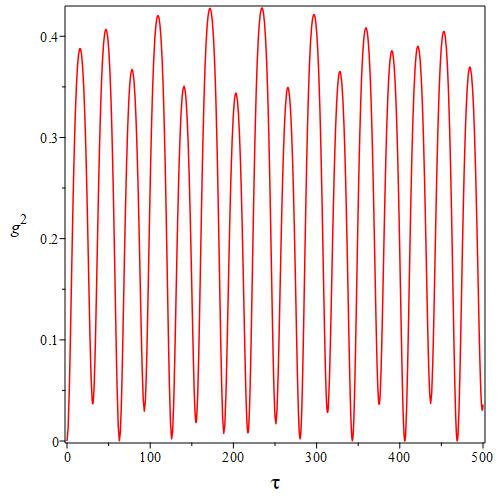} &
    \includegraphics[width=0.3\textwidth]{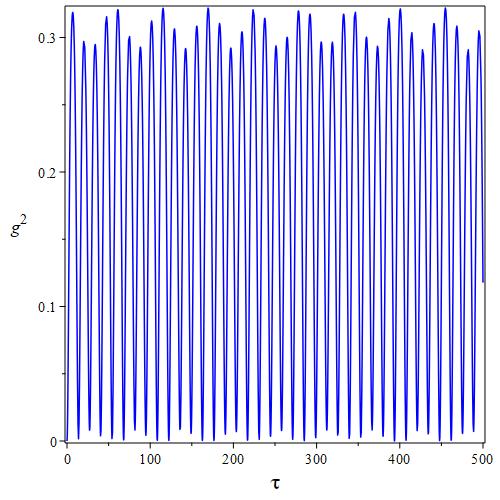} &
    \includegraphics[width=0.3\textwidth]{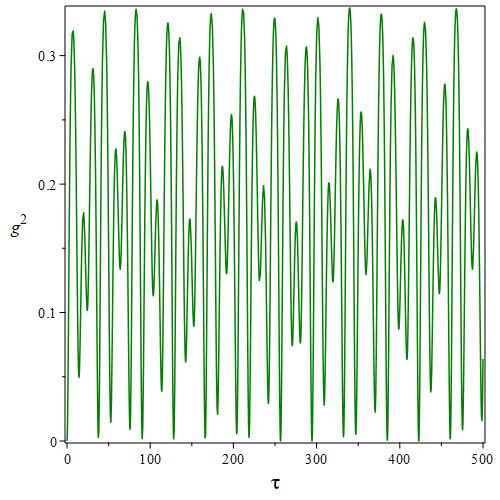} \\
     \text{a}\label{fig:sx1}&
    \text{b}\label{fig:sx2}&
    \text{c}\label{fig:sx2}\\
\end{tabular}
\caption{(Color online) Dynamics of the second order correlation $g^2$ of TLA with the same conditions as stated in Fig. (\ref{fig:inversion}).}
\label{co}
\end{table}
According to the Fig. (\ref{co}), Based on the numerical results, we illustrate the influence of the parameters $\Omega_e$, $\chi$ and $g_i$ on the statistical properties of the deformed field in the presence of a microwave field. It is clear that $g^2(0)<1$ indicates Sub-Poissonian statistics, and light is considered anti-bunched if $g^2(\tau)>g^2(0)$.
\section*{Quantum entanglement}\label{sec4}
\noindent In this section, we are attempting to analyses the quantum correlation between a V-type three-level atom and a single-mode coherent field by their entanglement and can be obtained using the subsystem entropy. According to the von Neumann entropy, the reduced field entropy is
defined as
\bea
S_F(t)=S_A(t)=-Tr_A[\hat{\rho}(t)ln(\hat{\rho}(t))]=-\sum_{j=1}^{3}\lambda_j ln(\lambda_j),
\eea
where the atom's reduced density matrix is given by,
\bea
\hat{\rho}_A(t)&&=Tr_{F}\left(|\psi(t)\rangle\langle \psi(t)|\right),\nonumber\\
&&=\begin{pmatrix}
                  c_1c^*_1 & 0 & 0 \\
                  0 & c_2c^*_2 & c_3c^*_2 \\
                  0 & c_2c^*_3 & c_3c^*_3
                \end{pmatrix},
\eea
and $\lambda_j's$ are the eigenvalues of $\rho_A(t)$. In Fig. (\ref{en}) the evolution of the von Neumann entropy in terms of the scaled time is depicted. The entropy increases from its minimum and then tends to exhibit a structure with rapid oscillations, there is a considerable amount of entanglement, as time evolves. From these results, we note that the deformation effect can enhance the amplitude of entropy oscillations and it is also seen that the minimum value of entropy is increased in the presence of microwave field.  On the other hand, when $f(\hat{n})=\sqrt{1+\chi\hat{n}^2}$, the evolution of the field entropy is quite interesting. This suggests that the Kerr-induced interaction has a detrimental impact on the amount of TLA-field entanglement as time significantly increases. The curve fluctuates between zero and one in an excited state. This means that the increase and presence of $\Omega_e, \chi$ and $g_i$ lead to a higher degree of entanglement between the atom and the field.
\begin{table}[ht!]
\centering
\begin{tabular}{ccc}
    \includegraphics[width=0.3\textwidth]{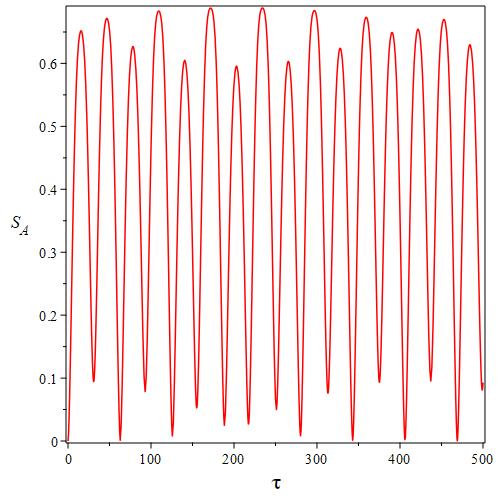} &
    \includegraphics[width=0.3\textwidth]{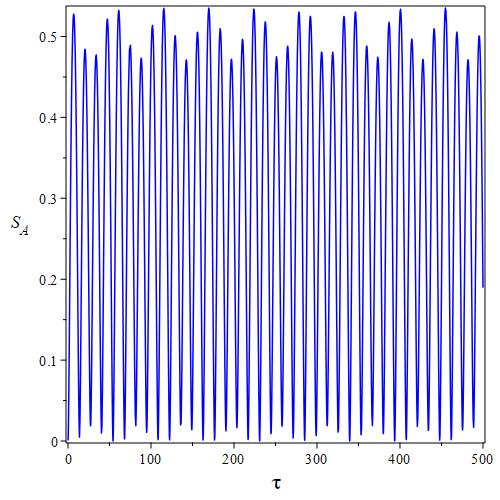} &
    \includegraphics[width=0.3\textwidth]{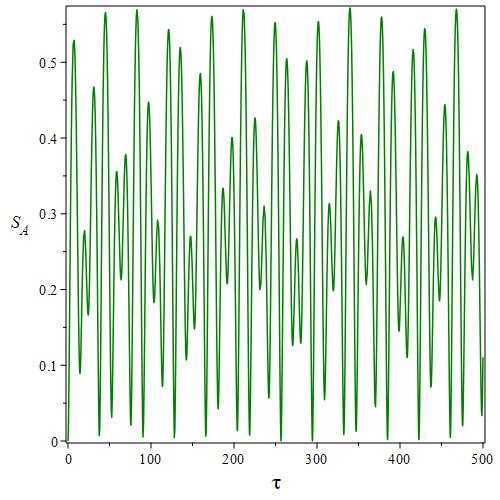} \\
     \text{a}\label{fig:sx1}&
    \text{b}\label{fig:sx2}&
    \text{c}\label{fig:sx2}\\
\end{tabular}
\caption{(Color online) Dynamics of the TLA Neumann entropy  $S_A(t)$ with the same conditions as stated in Fig. (\ref{fig:inversion}).}
\label{en}
\end{table}
\section*{Photon statistics}\label{sec5}
\noindent One of the most remarkable nonclassical effects of a quantum system is its sub-Poissonian photon statistics.
To determine such characteristics, we consider the Mandel $Q$ parameter defined by \cite{mandel1979sub}
\bea
Q=\frac{\langle(\hat{A}^\dag\hat{A})^2\rangle-\langle\hat{A}^\dag\hat{A}\rangle^2}{\langle\hat{A}^\dag\hat{A}\rangle}-1,
\eea
which measures the deviation of the photon number distribution from Poissonian statistics. Clearly, $Q$ turns zero for a coherent field and the minimal value $Q=-1$ is achieved  for Fock states, which have a well-defined number of photons by definition. If, $-1\leq Q<0$ the field statistics are Sub-Poissonian, and the phase space distribution cannot be interpreted as a classical probability function.
Therefore, the negativity of $Q$ is a sufficient condition for a field to be nonclassical. Again, if $Q>0$, the states exhibit  Super-Poissonian statistics, but no conclusions can be drawn about its  nature, for example, $Q$ parameter for a thermal field is always positive \cite{wang2023quantum}.
The Figs.~(\ref{Q}) indicate that the intensity-dependent coupling and the Rabi frequency leads to a completely negative $Q$ function, as well as altering  its chaotic behavior between positive and negative values to nearly regular oscillations in the negative range. Additionally, we observe that when $f(n)=\sqrt{1+\chi\hat{n}^2}$, increase and presence of $\Omega_e, \chi$ and $g_i$  add an irregularity in the behavior of the $Q$ function.
\begin{table}[ht!]
\centering
\begin{tabular}{ccc}
    \includegraphics[width=0.3\textwidth]{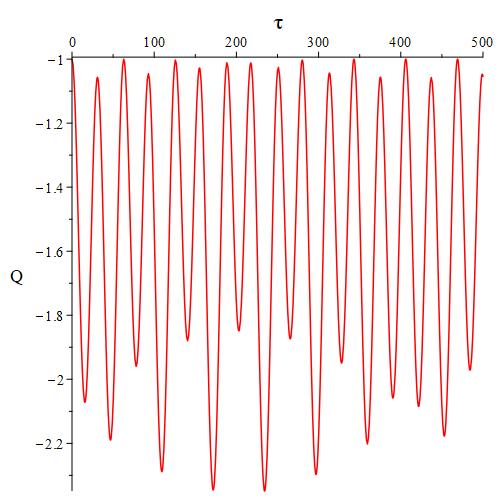} &
    \includegraphics[width=0.3\textwidth]{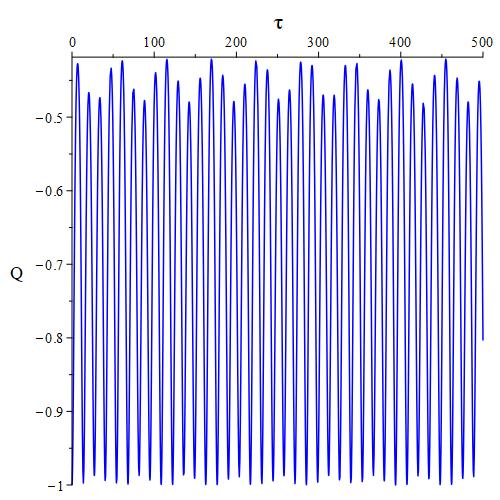} &
    \includegraphics[width=0.3\textwidth]{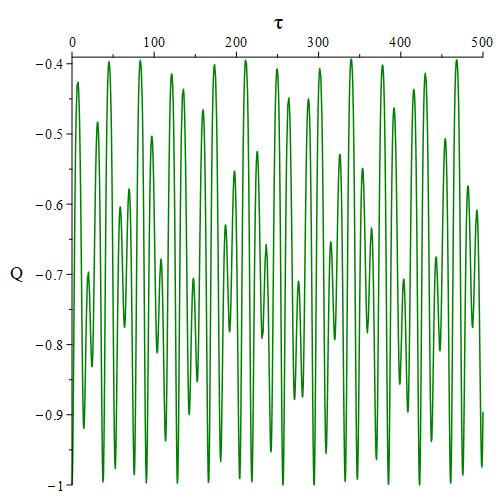} \\
     \text{a}&
    \text{b}&
    \text{c}\\
\end{tabular}
 \caption{(Color online) Mandel Q parameter with the same conditions as stated in Fig. (\ref{fig:inversion}).}\label{Q}
\end{table}
\subsection*{Husimi Q function}
\noindent The Husimi $Q$ function, a fundamental concept in quantum mechanics, serves as a crucial tool for characterizing the quantum state of a system in phase space. It is a quasi-probability distribution function that provides a unique perspective on the state of a quantum system by representing the probability amplitude in phase space rather than in the traditional position or momentum basis. Unlike the Wigner function, which may produce negative values, the Husimi $Q$ function is always positive, resembling a classical probability density function. It is a bounded function without any singularities. This characteristic makes it particularly useful for interpreting and visualizing the quantum state of a system in a more intuitive and classical-like manner.

Furthermore, the Husimi $Q$ function is defined as the projection of the quantum state onto coherent states, which are states with well-defined phase and amplitude properties. By taking the coherent state expectation value of the density matrix, the Husimi $Q$ function captures the essence of the quantum state's phase space distribution. Its bounded nature and lack of singularities make it a valuable mathematical tool for analyzing and understanding the dynamics of quantum systems. Overall, the Husimi $Q$ function bridges the gap between quantum and classical descriptions, providing a comprehensive and insightful representation of quantum states in a familiar phase space framework.
This function is specifically defined as \cite{glauber1963coherent}
\bea
Q(\beta,t)&=&\frac{1}{\pi}|\langle \beta|\psi(t)\rangle|^2\nn\\
&=&\frac{1}{\pi}\exp^{-|\beta|^2}\sum_{n=0}^{\infty}\frac{|\beta|^{2n}}{n!}\times
\Big[\Big(\frac{|\beta|^2}{n+1}\Big)|C_1(n+1,t)|^2+|C_2(n,t)|^2+|C_3(n,t)|^2\Big].
\eea
In Fig. (\ref{fig:husimi}), we have illustrated the mesh plots of the $Q$ function in the complex $\beta$-plane assuming $x = Re(\beta)$ and $y = Im(\beta)$.\\
This illustration highlights the presence of the Kerr-induced interaction($f(n)=\sqrt{1+\chi\hat{n}^2}$) and microwave field.
 We have identified a circular region with a radius centered at $x =0,y=0$, indicating that the intensity-dependent coupling and microwave field have no impact on the  $Q$ function.\\
Furthermore, we observe that the shape of the distribution function remains nearly unchanged when altering the scale time and considering the case non-zero (zero) $\chi$ parameter.
\begin{table}[ht!]
\centering
\begin{tabular}{ccc}
    \includegraphics[width=0.3\textwidth]{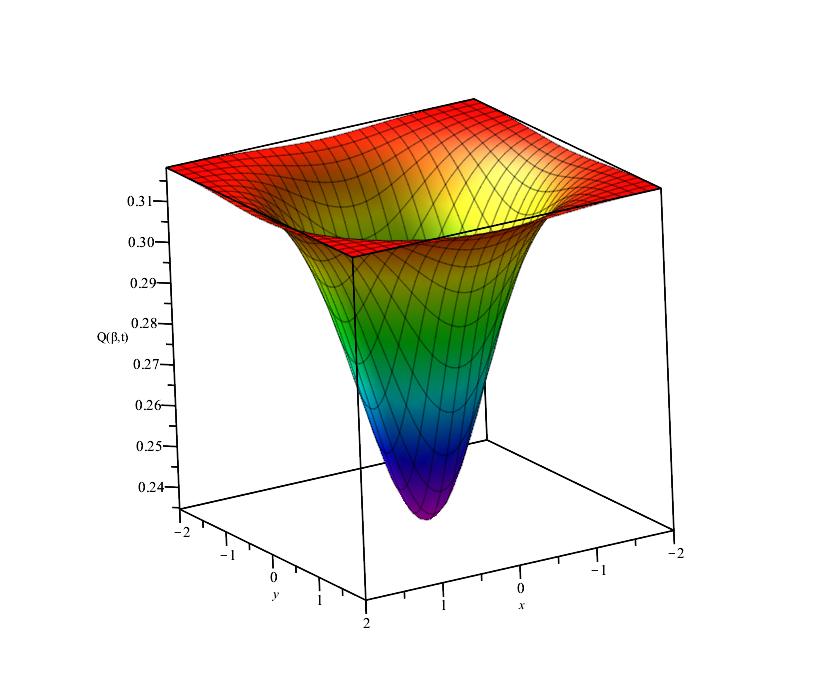} &
    \includegraphics[width=0.3\textwidth]{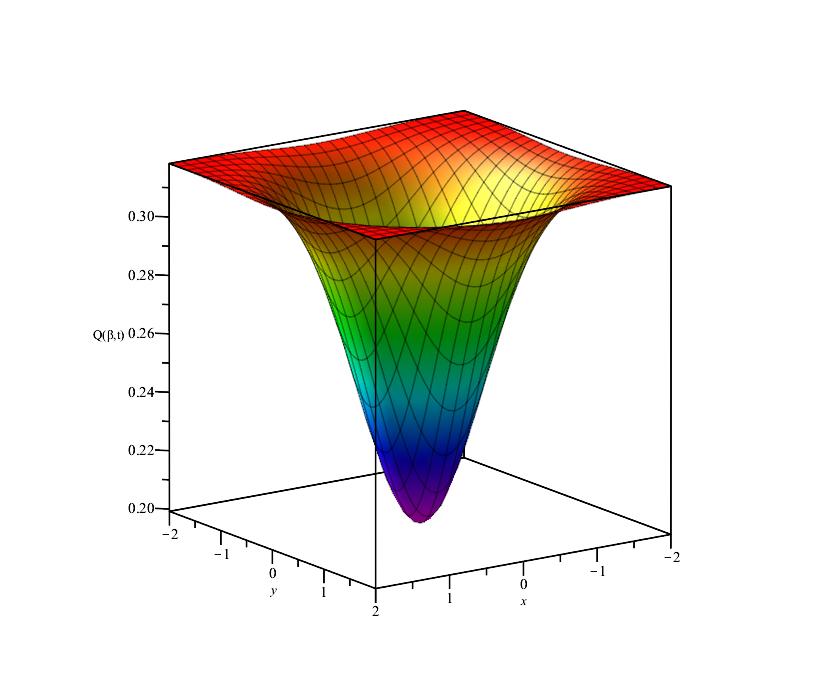} \\
     \text{$\chi=0$} &
    \text{$\chi=0.2$} \\
\end{tabular}
 \caption{(Color online) Husimi Q function with the same conditions as stated in Fig. (\ref{fig:inversion}}\label{fig:husimi}
\end{table}
\section*{Squeezing of different orders}
\noindent Squeezing of different orders is a concept primarily found in the field of quantum optics and quantum information science. Squeezing refers to the process by which the quantum noise in one variable (such as the amplitude or phase of light) is reduced below the standard quantum limit or compared to a coherent state, at the expense of increased noise in the conjugate variances adheres to the uncertainty principle. This reduction in noise is valuable for various applications, including precision measurements, quantum communication, optical communication networks, interferometric techniques, and optical waveguide taps. The phenomena of squeezing can be described through linear combinations of photon creation and annihilation operators, as well as higher-order combinations of these operators \cite{yuen1978optical, caves1985new, shapiro1980optical}.
\subsection*{Normal and Amplitude Squared squeezing}\label{sec6}
\noindent Normal squeezing, also known as first-order squeezing, refers to the reduction of quantum noise in one of the quadrature components of the electromagnetic field below the standard quantum limit, while the noise in the conjugate quadrature is increased to maintain the Heisenberg uncertainty principle. The quadrature components, $\hat{x}$ and $\hat{p}$, are analogous to position and momentum in quantum mechanics and can be expressed in terms of the photon creation and annihilation operators:
\bea
\hat{x}=\frac{1}{\sqrt{2}}(\hat{a}+\hat{a}^\dagger),\nonumber\\
\hat{p}=\frac{1}{\sqrt{2i}}(\hat{a}-\hat{a}^\dagger).
\eea
In a squeezed state, the variance of one quadrature, say $\hat{x}$, is reduced
\bea
\langle(\Delta\hat{x})^2\rangle<\frac{1}{2}.
\eea
This reduction is compensated by an increase in the variance of the conjugate quadrature, $\hat{p}$, such that
\bea
\langle(\Delta\hat{x})^2\rangle\langle(\Delta\hat{p})^2\rangle\geq\frac{1}{4}.
\eea
This relationship ensures that the Heisenberg uncertainty principle is satisfied. Normal squeezing is typically achieved using nonlinear optical processes such as parametric down-conversion or four-wave mixing, which allow for the manipulation of the quantum noise properties of the light field.
Normal squeezing has significant applications in precision measurements, such as in gravitational wave detectors like LIGO, where squeezed light is used to enhance sensitivity by reducing quantum noise \cite{yuen1978optical, caves1985new, shapiro1980optical}.
To derive a condition and a quantitative measure of First order squeezing, it is advantageous to introduce the following parameters
\bea
s^1_x&=&4(\Delta\hat{x})^2-1\nonumber\\
&=&2\langle \hat{A}^\dag\hat{A}\rangle+\langle \hat{A}^2\rangle+\langle \hat{A}^{\dag2}\rangle-\langle \hat{A}\rangle^2-\langle \hat{A}^\dag\rangle^2-2\langle \hat{A}\rangle\langle \hat{A}^\dag\rangle,
\eea
and
\bea
s^1_p&=&4(\Delta\hat{p})^2-1\nonumber\\
&=&2\langle \hat{A}^\dag\hat{A}\rangle-\langle \hat{A}^2\rangle-\langle \hat{A}^{\dag2}\rangle+\langle \hat{A}\rangle^2+\langle \hat{A}^\dag\rangle^2-2\langle \hat{A}\rangle\langle \hat{A}^\dag\rangle.
\eea
Amplitude squared squeezing refers to the reduction of quantum noise in the second-order moments of the amplitude quadratures of the electromagnetic field. This type of squeezing is more complex than first-order squeezing and involves higher-order correlations between photons.

In this context, consider the amplitude quadrature $\hat{X}=\frac{1}{\sqrt{2}}(\hat{A}+\hat{A}^\dagger)$ and the phase quadrature $\hat{P}=\frac{1}{\sqrt{2i}}(\hat{A}-\hat{A}^\dagger)$. Amplitude squared squeezing occurs if the variance is reduced below the corresponding value for a coherent state $\langle(\Delta\hat{X}^2)^2\rangle<1/4$ or $\langle(\Delta\hat{P}^2)^2\rangle<1/4$.
This type of squeezing is significant because it allows for the reduction of quantum noise in higher-order moments, which is beneficial for various advanced quantum information processing tasks, such as quantum state engineering and quantum metrology.
Amplitude squared squeezing can be realized through nonlinear optical processes, such as second-harmonic generation or interactions in nonlinear crystals, where the higher-order correlations between photons are manipulated to achieve the desired noise reduction.
This involves the reduction of quantum noise in the second-order moments of the amplitude and phase quadratures, extending beyond the usual first-order squeezing to higher-order correlations \cite{hillery1987amplitude}.
The second-order squeezing occurs if the parameters
\bea
s^2_x=2\langle( \hat{A}^\dag\hat{A})^2\rangle-2\langle \hat{A}^\dag\hat{A}\rangle+\langle \hat{A}^4\rangle+\langle \hat{A}^{\dag4}\rangle-\langle \hat{A}^2\rangle^2-\langle \hat{A}^{\dag2}\rangle^2-2\langle \hat{A}^2\rangle\langle \hat{A}^{\dag2}\rangle,
\eea
and
\bea
s^2_p=2\langle( \hat{A}^\dag\hat{A})^2\rangle-2\langle \hat{A}^\dag\hat{A}\rangle-\langle \hat{A}^4\rangle-\langle \hat{A}^{\dag4}\rangle+\langle \hat{A}^2\rangle^2+\langle \hat{A}^{\dag2}\rangle^2-2\langle \hat{A}^2\rangle\langle \hat{A}^{\dag2}\rangle.
\eea
Considering the Eq.~(\ref{si}), we understand that the expectation values of $\langle \hat{A}^4\rangle, \langle \hat{A}^{\dag4}\rangle, \langle \hat{A}^{\dag2}\rangle$ and $\langle \hat{A}^4\rangle$ will be zero.
\begin{table}[ht!]
\centering
\begin{tabular}{ccc}
    \includegraphics[width=0.3\textwidth]{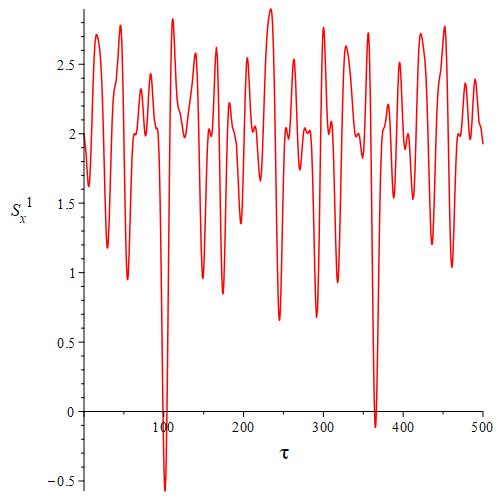} &
    \includegraphics[width=0.3\textwidth]{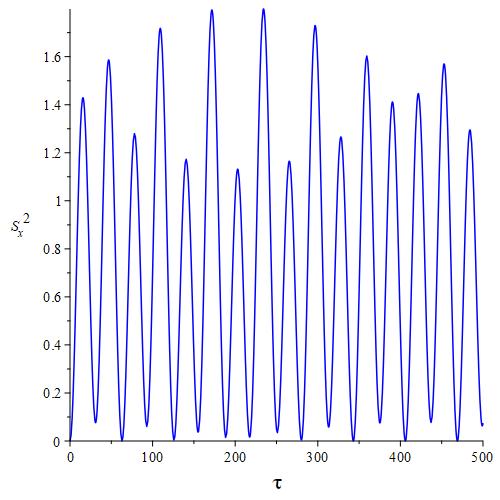}\\
    \text{a}\label{sx1}&
    \text{b}\label{sx2}\\
      \includegraphics[width=0.3\textwidth]{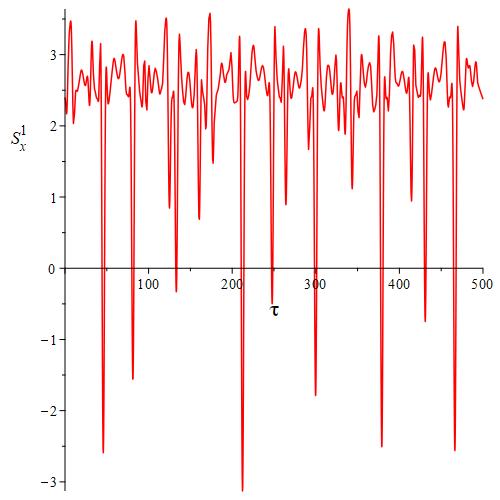}&
    \includegraphics[width=0.3\textwidth]{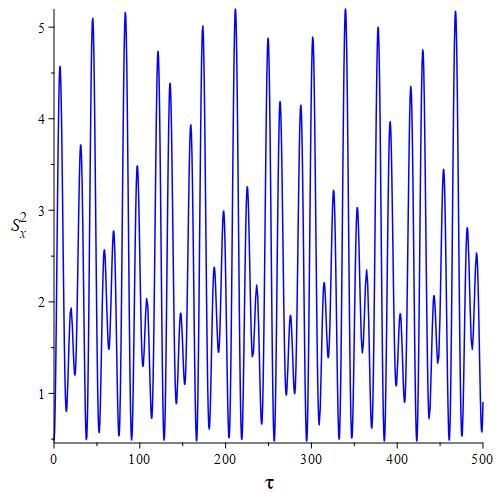}\\
     \text{c}\label{sssx1}&
    \text{d}\label{sssx2}\\
\end{tabular}
 \caption{(Color online) Dynamics of the Normal and Amplitude squared squeezing of TLA for the parameters $\Omega=0.2$, $\omega_3=0.5, \omega_2=0.4, \omega_1=0.3$, $n=1$, the intensity-dependent coupling $f(\hat{n})=\sqrt{1+\chi\hat{n}^2}$ and the $\tau=\Omega t$.
  The upper row. (a,b) corresponds  to the $\Omega_e=0.04, g_1=0.04, g_2=0.06$ and $\chi=0$, and the bottom row. (c,d) corresponds to the $\Omega_e=0.08, g_1=0.06, g_2=0.08$ and $\chi=0.2$.}\label{sx}
\end{table}
Based on the First order squeezing in  Fig.~(\ref {sx}a), we observe that in the absence of intensity-dependent coupling  $(f(\hat{n})=1)$ and with Low values of Rabi frequency and coupling parameters, squeezing occurs only during revival times. However, in the second-order squeezing depicted in  Fig.~(\ref {sx}b), we see that under similar conditions, squeezing does not occur, but the range of oscillations increases. Now, in the presence of these parameters, as shown in Figs.~(\ref {sx}c, \ref {sx}d), it is observed that the volume of the first-order squeezing increases and the depth of squeezing becomes significantly greater than the first-order squeezing, with an increase in the number of oscillations. In the amplitude squared squeezing case, squeezing does not occur, but the range of oscillations increases.
\section*{CONCLUSION}\label{con}
\noindent In this paper, we have investigated a quantum system comprising a nonlinear generalization of a V-type three-level atom interacting with a deformed single-mode field, simultaneously driven by a cavity photon and a microwave field. Using the framework of f-deformed algebra and selecting an appropriate nonlinear function $f(\hat{n})=\sqrt{1+\chi\hat{n}^2}$ corresponding to a Kerr constant, we proposed a nonlinear Hamiltonian to describe a system with two types of interactions: the nonlinear coupling effect of the atom and deformed radiation field, and the driving effect of the external microwave field. By solving the corresponding Schr\"{o}dinger equation, we derived the state vector for the entire system at any time, with specific initial conditions assigned to the atom and the field.

We systematically examined the effects of intensity-dependent coupling and microwave field on various quantum measures such as population inversion, degree of quantum entanglement, normal and higher-order squeezing. Quantum entanglement was evaluated computationally using von-Neumann entropy. The well-known probability and quasi-probability distribution functions, the Husimi-Q and Wigner functions, were also derived.
Our initial calculations revealed that the time evolution of the population strongly depends on the atom-cavity coupling coefficient
$g$ and the Rabi frequency $(\Omega)$, which reflects the intensity of the external microwave field. When an external microwave field is introduced to induce coupling between the two upper levels, population exchange between these levels occurs, leading to a series of intriguing phenomena. The population dynamics of the two upper levels no longer follow the same pattern over time due to the driving effect of the external field, resulting in more population transferring to the highest level in each cycle. The overall system dynamics under the combined influence of the deformed cavity photon and the external field show a superposition of fast and slow periodic oscillations, where the fast oscillation is due to the deformed cavity photon, and the slow oscillation is driven by the microwave field. The frequency of the slow oscillation increases gradually with the intensity of the external field. The microwave field plays a more dominant role when the Rabi frequency and $\chi$ parameters are smaller, while the cavity photon effect is more pronounced for larger parameters.

The results indicated that the intensity-dependent coupling completely removes the super-Poissonian character of the atom-field state, although its depth of negativity is reduced. For linear coupling $f(\hat{n})=1$, Rabi frequency and coupling parameters have minimal effect on the Mandel parameter, while for nonlinear coupling, they cause negligible perturbations to its periodic behavior. When the atom-field coupling is independent of the field intensity, squeezing is observed for a short interval, and the Husimi-Q function shows no response to changes in atom-field coupling and other constraints. The field entropy oscillates between zero and its maximum value over time, with linear coupling causing distortions in its behavior unless the system is in resonance. Intensity-dependent coupling regularizes the entropy oscillations except when detunings are non-zero.
\bibliography{ref}
\end{document}